\documentstyle[aps,epsf]{revtex}

\begin{document}

\title{Anisotropic Relativistic Stellar Models}
\author{T. Harko\footnote{E-mail: tcharko@hkusua.hku.hk}}
\address{Department of Physics, The University of Hong Kong,
Pokfulam Road, Hong Kong, P. R. China.}
\author{M. K. Mak\footnote{E-mail:mkmak@vtc.edu.hk}}
\address{Department of Physics, The Hong Kong University of Science and Technology,Clear Water Bay, Hong Kong, P. R. China.}
% \date{July 14, 2001}
\maketitle

\begin{abstract}

We present a class of exact solutions of Einstein's gravitational field equations
describing spherically symmetric and static anisotropic stellar type configurations.
The solutions are obtained by assuming a particular form of the anisotropy factor. The
energy density and both radial and tangential pressures are finite and positive inside
the anisotropic star. Numerical results show that the basic physical parameters (mass and radius)
of the model can describe realistic astrophysical objects like neutron stars.



\end{abstract}

%\narrowtext

\section{Introduction}

Since the pioneering work of Bowers and Liang \cite{BoLi74} there is an extensive
literature devoted to the study of anisotropic spherically symmetric static general
relativistic configurations. The study of static anisotropic fluid spheres is important
for relativistic astrophysics. The theoretical investigations of Ruderman \cite{Ru72}
about more realistic stellar models show that the nuclear matter may be anisotropic at least in
certain very high density ranges ($\rho >10^{15}g/cm^{3}$), where the nuclear interactions
must be treated relativistically. According to these views in such massive stellar
objects the radial pressure may not be equal to the tangential one. No celestial body
is composed of purely perfect fluid. Anisotropy in fluid pressure could be introduced
by the existence of a solid core or by the presence of type $3A$ superfluid \cite{KiWe90}, different
kinds of phase transitions \cite{So80}, pion condensation \cite{Sa72} or by other
physical phenomena. On the scale of galaxies, Binney and Tremaine \cite{BiTr87} have
considered anisotropies in spherical galaxies, from a purely Newtonian point of view. Other
source of anisotropy, due to the effects of the slow rotation in a star, has been
proposed recently by Herrera and Santos \cite{HeSa95}.

The mixture of two gases (e.g., monatomic hydrogen, or ionized hydrogen and electrons)
can formally be also described as an anisotropic fluid \cite{Le80}. More generally,
 when the fluid is composed of two fluids the total energy-momentum tensor is
\begin{equation}\label{1}
T^{ik}=\left( P_{1}+\rho _{1}\right) U^{i}U^{k}-P_{1}g^{ik}+\left(
P_{2}+\rho _{2}\right) W^{i}W^{k}-P_{2}g^{ik},
\end{equation}
where $U_{i}U^{i}=1$ and $W_{i}W^{i}=1$. By means of the transformations
\begin{equation}\label{2}
U^{\ast i}=U^{i}\cos \alpha +\sqrt{\frac{P_{2}+\rho _{2}}{P_{1}+\rho _{1}}}%
W^{i}\sin \alpha ,W^{\ast i}=-\sqrt{\frac{P_{2}+\rho _{2}}{P_{1}+\rho _{1}}}%
U^{i}\sin \alpha +W^{i}\cos \alpha , 
\end{equation}
the energy momentum tensor (\ref{1}) can always be cast into the standard
form for anisotropic fluids,
\begin{equation}\label{3}
T^{ik}=\left( \rho +p\right) V^{i}V^{k}-pg^{ik}+\left( \sigma -p\right) \chi
^{i}\chi ^{k},  
\end{equation}
where $V^{i}=U^{\ast i}/\sqrt{U^{\ast i}U_{i}^{\ast }}$, $\chi ^{i}=W^{\ast
i}/\sqrt{-W^{\ast i}W_{i}^{\ast }}$, $\rho =T_{ik}V^{i}V^{k}$, $%
p=P_{1}+P_{2} $ and $\sigma $ is a complicated function of the densities and
pressures of the two fluids \cite{Ba82}.

The starting point in the study of fluid spheres is represented by the
interior Schwarzschild solution from which all problems involving spherical
symmetry can be modeled. Bowers and Liang \cite{BoLi74} have investigated
the possible importance of locally anisotropic equations of state for
relativistic fluid spheres by generalizing the equations of hydrostatic
equilibrium to include the effects of local anisotropy. Their study shows
that anisotropy may have non-negligible effects on such parameters as
maximum equilibrium mass and surface redshift. Heintzmann and Hillebrandt 
\cite{HeHi75} studied fully relativistic, anisotropic neutron star models at
high densities by means of several simple assumptions and have shown that
for arbitrary large anisotropy there is no limiting mass for neutron stars,
but the maximum mass of a neutron star still lies beyond $3-4M_{\odot}$.
Hillebrandt and Steinmetz \cite{HiSt76} considered the problem of stability
of fully relativistic anisotropic neutron star models. They derived the
differential equation for radial pulsations and showed that there exists a
static stability criterion similar to the one obtained for isotropic
models. Anisotropic fluid sphere configurations have been analyzed, using
various Ansatze, in \cite{Ba82} and \cite{CoHeEsWi81}-\cite{HaMa00}. For
static spheres in which the tangential pressure differs from the radial one,
Bondi \cite{Bo92} has studied the link between the surface value of the
potential and the highest occurring ratio of the pressure tensor to the local
density. Chan, Herrera and Santos \cite{ChHeSa93} studied in detail the role
played by the local pressure anisotropy in the onset of instabilities and
they showed that small anisotropies might in principle drastically change
the stability of the system. Herrera and Santos \cite{HeSa95} have extended
the Jeans instability criterion in Newtonian gravity to systems with
anisotropic pressures. Recent reviews on isotropic and anisotropic fluid
spheres can be found in \cite{DeLa98}-\cite{HeSa97}. There are very few interior
solutions (both isotropic and anisotropic) of the gravitational field equations satisfying the required
general physical conditions inside the star. From $127$ published solutions analyzed in
\cite{DeLa98} only $16$ satisfy all the conditions. 

In the present paper we consider a class of exact solutions of the gravitational
field equations for an anisotropic fluid sphere, corresponding to a specific
choice of the anisotropy parameter. The metric functions can be represented in a closed form 
in terms of elementary functions. In the isotropic limit we recover the interior
solutions previously found first by Buchdahl \cite{Bu59} and then by Durgapal and Bannerji \cite{DuBa83}.
Hence our solution can be considered the generalization to the anisotropic case of these solutions. 
All the physical parameters like the energy density, pressure and metric tensor components
are regular inside the anisotropic star, with the speed of sound less than the speed
of light. Therefore this solution can give a satisfactory description of realistic
astrophysical compact objects like neutron stars. Some explicit numerical models of relativistic
anisotropic stars, with a possible astrophysical relevance, are also presented.

This paper is organized as follows. In Section 2 we present an exact class of solutions for
an anisotropic fluid sphere. In Section 3 we present neutron star models with
possible astrophysical relevance. The results are summarized and discussed
in Section 4.

\section{Non-Singular Models for Anisotropic Stars}

In standard coordinates $x^{i}=\left( t,r,\theta ,\phi \right) $, the
general line element for a static spherically symmetric space-time takes the
form
\begin{equation}\label{4}
ds^{2}=A^{2}(r)dt^{2}-V^{-1}(r)dr^{2}-r^{2}\left( d\theta ^{2}+\sin
^{2}\theta d\phi ^{2}\right).  
\end{equation}

Einstein's gravitational field equations are (where natural units $8\pi
G=c=1$ have been used throughout):
\begin{equation}\label{5}
R_{i}^{k}-\frac{1}{2}R\delta _{i}^{k}=T_{i}^{k}.  
\end{equation}

For an anisotropic spherically symmetric matter distribution the components
of the energy-momentum tensor are of the form
\begin{equation}\label{6}
T_{i}^{k}=\left( \rho +p_{\perp }\right) u_{i}u^{k}-p_{\perp }\delta
_{i}^{k}+\left( p_{r}-p_{\perp }\right) \chi _{i}\chi ^{k},  
\end{equation}
where $u^{i}$ is the four-velocity $Au^{i}=\delta _{0}^{i}$, $\chi ^{i}$ is
the unit spacelike vector in the radial direction $\chi ^{i}=\sqrt{V}\delta
_{1}^{i}$, $\rho $ is the energy density, $p_{r}$ is the pressure in the
direction of $\chi ^{i}$ (normal pressure) and $p_{\perp }$ is the pressure
orthogonal to $\chi _{i}$ (transversal pressure). We assume $p_{r}\neq
p_{\perp }$. The case $p_{r}=p_{\perp }$ corresponds to the isotropic fluid
sphere. $\Delta =p_{\perp }-p_{r}$ is a measure of the anisotropy and is
called the anisotropy factor \cite{HeLe85}.

A term $\frac{2\left( p_{\perp
}-p_{r}\right) }{r}$ appears in the conservation equations $T_{k;i}^{i}=0$,
(where a semicolon $;$ denotes the covariant derivative with respect to the metric),
representing a force that is due to the anisotropic nature of the fluid.
This force is directed outward when $p_{\perp }>p_{r}$ and inward when $%
p_{\perp }<p_{r}$. The existence of a repulsive force (in the case $p_{\perp
}>p_{r}$) allows the construction of more compact objects when using
anisotropic fluid than when using isotropic fluid \cite{GoMe94}.

For the metric (\ref{4}) the gravitational field equations (\ref{5}) become
\begin{equation}\label{7}
\rho =\frac{1-V}{r^{2}}-\frac{V^{\prime }}{r},p_{r}=\frac{2A^{\prime }V}{Ar}+%
\frac{V-1}{r^{2}},  
\end{equation}
\begin{equation}\label{8}
V^{\prime }\left( \frac{A^{\prime }}{A}+\frac{1}{r}\right) +2V\left( \frac{%
A^{\prime \prime }}{A}-\frac{A^{\prime }}{rA}-\frac{1}{r^{2}}\right)
=2\left( \Delta -\frac{1}{r^{2}}\right),  
\end{equation}
where $\prime =\frac{d}{dr}$.

It is convenient to introduce the following substitutions \cite{Bu59}:
\begin{equation}\label{9}
V=1-2x\eta ,x=r^{2},\eta \left( r\right) =\frac{m(r)}{r^{3}},m(r)=\frac{1}{2}%
\int_{0}^{r}\xi ^{2}\rho \left( \xi \right) d\xi .  
\end{equation}

$m(r)$ represents the total mass content of the distribution within the
fluid sphere of radius $r$. Hence, we can express Eq. (\ref{8}) in the form
\begin{equation}\label{10}
\left( 1-2x\eta \right) \frac{d^{2}A}{dx^{2}}-\left( x\frac{d\eta }{dx}+\eta
\right) \frac{dA}{dx}-\left( \frac{1}{2}\frac{d\eta }{dx}+\frac{\Delta }{4x}%
\right) A=0. 
\end{equation}

For any physically acceptable stellar models, we require the condition that
the energy density is positive and finite at all points inside the fluid
spheres. In order to have a monotonic decreasing energy density $\rho =\frac{%
2}{r^{2}}\frac{d}{dr}\left( \eta r^{3}\right) $ inside the star we chose the
function $\eta $ in the form
\begin{equation}\label{11}
\eta =\frac{a_{0}}{2\left( 1+\psi \right) },  
\end{equation}
where $\psi =c_{0}x$ and $a_{0},c_{0}$ are non-negative constants. We
introduce a new variable $\lambda $ by means of the transformation
\begin{equation}\label{12}
\lambda =\frac{\left( a_{0}-c_{0}\right) \left( 1+\psi \right) }{a_{0}}=%
\frac{1}{3}\left( 1-\Delta _{0}\right) \left( 1+\psi \right) .  
\end{equation}

We also chose the anisotropy parameter as
\begin{equation}\label{13}
\Delta =\frac{3c_{0}\Delta _{0}\psi }{\left( 2+\Delta _{0}\right) \left(
\psi +1\right) ^{2}},  
\end{equation}
where $\Delta _{0}=\frac{3c_{0}}{a_{0}}-2$. In the following we assume $%
\Delta _{0}\geq 0$, with $\Delta _{0}=0$ corresponding to the isotropic
limit. Hence $\Delta _{0}$ can be considered, in the present model, as a measure
of the anisotropy of the pressure distribution inside the fluid sphere. At the
center of the fluid sphere the anisotropy vanishes, $\Delta (0)=0$. For small
values of $r$, near the center, $\Delta (r)$ is an increasing function of $r$, but,
after reaching a maximum, the anisotropy decreases becoming negligible small at the
vacuum boundary of the star. 

Therefore with this choice of $\Delta $, Eq. (\ref{10}) becomes a hypergeometric equation,
\begin{equation}\label{14}
\lambda \left( \lambda -1\right) \frac{d^{2}A}{d\lambda ^{2}}+\frac{1}{2}%
\frac{dA}{d\lambda }-\frac{3}{4} A=0. 
\end{equation}

In the isotropic case $\Delta _{0}=0$ we obtain $\frac{a_{0}}{c_{0}}=\frac{3}{2}$%
. Consequently, in this limit we recover the results obtained by Buchdahl \cite{Bu59} and
Durgapal and Bannerji \cite{DuBa83}.

On integration we obtain the general solution of Eq. (\ref{14}) and the
general solution of the gravitational field equations for a static
anisotropic fluid sphere, with anisotropy parameter given by Eq. (\ref{13}),
in the following form, expressed in elementary functions:
\begin{equation}\label{15}
A=\sqrt{\alpha _{1}}\left\{ \left( 1+\psi \right) ^{3/2}+\beta _{1}\left[
5-2\Delta _{0}+2\left( 1-\Delta _{0}\right) \psi \right] \sqrt{2-\psi
+\Delta _{0}\left( 1+\psi \right) }\right\},  
\end{equation}
\begin{equation}\label{16}
V=1-\frac{3\psi }{\left( 2+\Delta _{0}\right) \left( 1+\psi \right) },
\end{equation}
\begin{equation}\label{17}
\rho =\frac{3c_{0}\left( 3+\psi \right) }{\left( 2+\Delta _{0}\right) \left(
1+\psi \right) ^{2}},p_{r}=\frac{3c_{0}\left( p_{2}-p_{1}\right) }{p_{3}}%
,p_{\perp }=p_{r}+\frac{3\Delta _{0}c_{0}\psi }{\left( 2+\Delta _{0}\right)
\left( 1+\psi \right) ^{2}},  
\end{equation}
\begin{equation}\label{18}
\frac{dp_{r}}{d\psi }=3c_{0}\left[ \frac{1}{p_{3}}\frac{d\left(
p_{2}-p_{1}\right) }{d\psi }+\left( p_{2}-p_{1}\right) \frac{d}{d\psi }%
\left( \frac{1}{p_{3}}\right) \right],  
\end{equation}
\begin{equation}\label{19}
\frac{dp_{\perp }}{d\psi }=\frac{dp_{r}}{d\psi }+\frac{3\Delta
_{0}c_{0}\left( 1-\psi \right) }{\left( 2+\Delta _{0}\right) \left( 1+\psi
\right) ^{3}},  
\end{equation}
where $\alpha _{1}$ and $\beta _{1}$ are constants of integration and we
denoted
\begin{equation}\label{20}
F\left( \psi \right) =\sqrt{2-\psi +\Delta _{0}\left( 1+\psi \right) },
\end{equation}
\begin{equation}\label{21}
p_{1}=\sqrt{1+\psi }\left[ 3\left( \psi -1\right) -2\Delta _{0}\left( 1+\psi
\right) \right] F\left( \psi \right),  
\end{equation}
\begin{equation}\label{22}
p_{2}=\beta _{1}\left[ 3\left( 2\psi +1\right) \left( \psi -2\right)
-4\Delta _{0}^{3}\left( 1+\psi \right) ^{2}+2\Delta _{0}^{2}\left( 7\psi
^{2}+5\psi -2\right) -\Delta _{0}\left( 16\psi ^{2}-7\psi -5\right) \right],
\end{equation}
\begin{equation}\label{23}
p_{3}=\left( \Delta _{0}+2\right) \left( 1+\psi \right) \left\{ \left(
1+\psi \right) ^{3/2}+\beta _{1}\left[ 5+2\psi -2\Delta _{0}\left( 1+\psi
\right) \right] F\left( \psi \right) \right\} F\left( \psi \right).
\end{equation}

In order to be physically meaningful, the interior solution for static fluid
spheres of Einstein's gravitational field equations must satisfy some general
physical requirements. The following conditions have been generally
recognized to be crucial for anisotropic fluid spheres\cite{HeSa97}:

a) the density $\rho $ and pressure $p_{r}$ should be positive inside the
star;

b) the gradients $\frac{d\rho }{dr}$, $\frac{dp_{r}}{dr}$ and $\frac{%
dp_{\perp }}{dr}$ should be negative;

c) inside the static configuration the speed of sound should be less than
the speed of light, i.e. $0\leq \frac{dp_{r}}{d\rho }\leq 1$ and $0\leq 
\frac{dp_{\perp }}{d\rho }\leq 1$;

d) a physically reasonable energy-momentum tensor has to obey the conditions 
$\rho \geq p_{r}+2p_{\perp }$ and $\rho +p_{r}+2p_{\perp }\geq 0$;

e) the interior metric should be joined continuously with the exterior
Schwarzschild metric, that is $A^{2}(a)=1-2u$, where $u=M/a$, $M$ is the
mass of the sphere as measured by its external gravitational field and $a$ is the
boundary of the sphere;

f) the radial pressure $p_{r}$ must vanish but the tangential pressure $%
p_{\perp }$ may not vanish at the boundary $r=a$ of the sphere. However, the
radial pressure is equal to the tangential pressure at the center of the
fluid sphere. 

By matching  Eq. (\ref{16}) on the boundary of the anisotropic sphere we
obtain
\begin{equation}\label{24}
V(a)=1-\frac{3c_{0}a^{2}}{\left( 2+\Delta _{0}\right) \left(
1+c_{0}a^{2}\right) }=1-2u_{anis}.  
\end{equation}

For the isotropic case, that is for $\Delta _{0}=0$, it is easy to show that 
$\frac{3c_{0}a^{2}}{4\left( 1+c_{0}a^{2}\right) }=u_{iso}$. Therefore the
mass-radius ratios for the anisotropic and isotropic spheres are related in the
present model by
\begin{equation}\label{25}
u_{anis}=\frac{2}{2+\Delta _{0}}u_{iso}.  
\end{equation}

Hence, the constants $\alpha _{1}$, $\beta _{1}$ and $c_{0}$ appearing in
the solution can be evaluated from the boundary conditions. Thus, by
denoting $X=c_{0}a^{2}$ we obtain
\begin{equation}\label{26}
X=c_{0}a^{2}=\frac{4u_{iso}}{3-4u_{iso}}=\frac{2u_{anis}\left( 2+\Delta
_{0}\right) }{3-4u_{anis}-2u_{anis}\Delta _{0}},  
\end{equation}
\begin{equation}
\alpha _{1}=\left( 1-2u_{anis}\right) \left\{ \left( 1+X\right) ^{3/2}+\beta
_{1}\left[ 5-2\Delta _{0}+2\left( 1-\Delta _{0}\right) X\right] \sqrt{%
2-X+\Delta _{0}\left( 1+X\right) }\right\} ^{-2},
\end{equation}
\begin{equation}\label{27}
\beta _{1}=\frac{\sqrt{\left( 1+X\right) \left[ 2-X+\Delta _{0}\left(
1+X\right) \right] }\left[ 3\left( X-1\right) -2\Delta _{0}\left( 1+X\right) %
\right] }{3\left( 2X+1\right) \left( X-2\right) -4\Delta _{0}^{3}\left(
1+X\right) ^{2}+2\Delta _{0}^{2}\left( 7X^{2}+5X-2\right) -\Delta _{0}\left(
16X^{2}-7X-5\right) }.  
\end{equation}

Note that equation (16) in \cite{DuBa83} has been amended as the  Eq. (\ref
{27}) presented here.

In order to find a general constraint for the anisotropy parameter $\Delta
_{0}$, we shall consider that the conditions $\rho _{0}=\rho \left( 0\right)
\geq 0$, $p_{0}=p(0)\geq 0$ and $\rho _{0}\geq 3p_{0}$ hold at the center of
the fluid sphere. Subsequently the parameters $\beta _{1}$ and $\Delta _{0}$
should be restricted to obey the following conditions:
\begin{equation}\label{28}
0\leq L\left( u_{iso},\Delta _{0}\right) =\frac{3+2\Delta _{0}-\beta
_{1}\left( 4\Delta _{0}^{2}-4\Delta _{0}+3\right) \sqrt{2+\Delta _{0}}}{%
1+\beta _{1}\left( 5-2\Delta _{0}\right) \sqrt{2+\Delta _{0}}}\leq 1.
\end{equation}

The general behavior of the function $L\left( u_{iso},\Delta _{0}\right) $
is represented in Fig. 1.

\newpage

\begin{figure}
\epsfxsize=10cm
\centerline{\epsffile{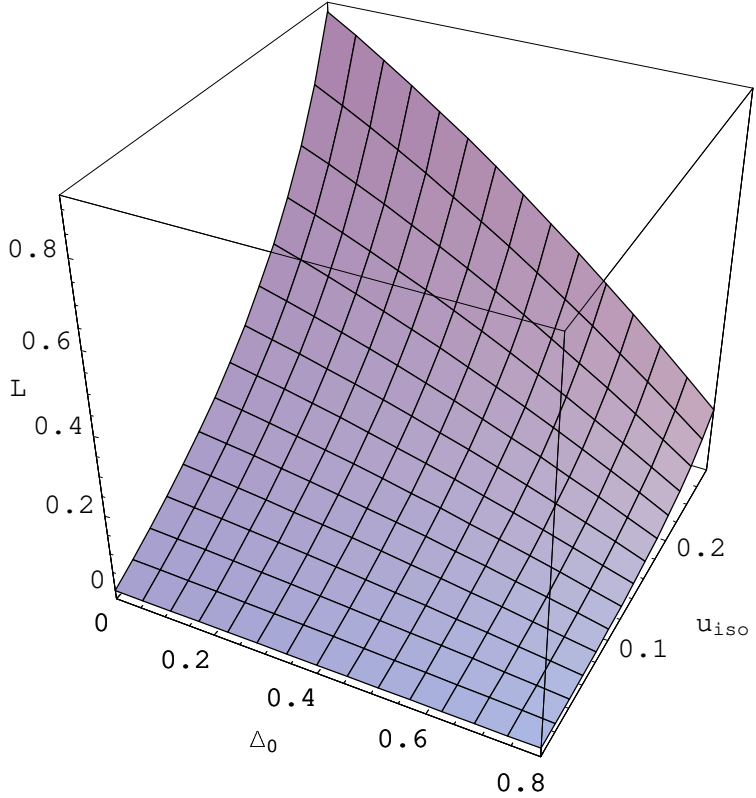}}
\caption{Variation of the function $L\left( u_{iso},\Delta _{0}\right) $
against the anisotropy parameter $\Delta _{0}$ and $u_{iso}$.}
\label{FIG1}
\end{figure}

Generally, the condition (\ref{28}) is satisfied for $0\leq \Delta _{0}<1$ and $u_{iso}\leq 0.3$.

\section{Astrophysical Applications}

When the thermonuclear sources of energy in its interior are exhausted, a
spherical star begins to collapse under the influence of gravitational
interaction of its matter content. The mass energy continues to increase and
the star ends up as a compact relativistic cosmic object such as neutron
star, strange star or black hole. Important observational quantities for
such objects are the surface redshift, the central redshift and the mass and
radius of the star.

For a relativistic anisotropic star described by the solution presented in
the previous Section the surface redshift $z_{s}$ is given by
\begin{equation}\label{29}
z_{s}=\left( 1-\frac{4}{2+\Delta _{0}}u_{iso}\right) ^{-1/2}-1.  
\end{equation}

The surface redshift is decreasing with increasing $\Delta _{0}$. Hence, at least
in principle, the study of redshift of light emitted at the surface
of compact objects can lead to the possibility of observational detection of
anisotropies in the internal pressure distribution of relativistic stars. Using the density variation parameter $\mu =\rho
(R)/\rho (0)$, Patel and Mehra \cite{PaMe95} discussed numerical estimates
of various physical parameters in their model and concluded that the surface
redshift in the isotropic case is greater than the surface redshift in the
anisotropic case. Hence, our results are very similar to that of \cite
{PaMe95}.

The central redshift $z_{c}$ is of the form
\begin{equation}\label{30}
z_{c}=\alpha _{1}^{-1/2}\left[ 1+\beta _{1}\left( 5-2\Delta _{0}\right) 
\sqrt{2+\Delta _{0}}\right] ^{-1}-1.  
\end{equation}

The variation of the central redshift of the neutron star against the
anisotropy parameter is represented in Fig. 2.

\newpage

\begin{figure}[h]
\epsfxsize=10cm
\centerline{\epsffile{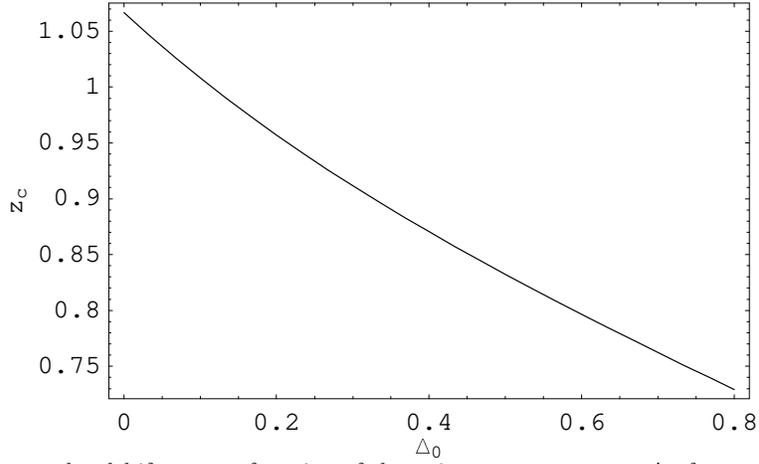}}
\caption{Variation of the central redshift $z_{c}$ as a function of the anisotropy
parameter $\Delta _{0}$ for a static anisotropic fluid sphere with 
$u_{iso}=0.27$.}
\label{FIG2}
\end{figure}

Clearly, in our model the anisotropy introduced in the pressure gives rise to a decrease in the
central redshift. Hence, as functions of the anisotropy, the central and surface redshifts have the same behavior.  

The stellar model presented here can be used to describe the interior
structure of the realistic neutron star. Taking the surface density of the
star as $\rho _{s}=2\times 10^{14}g/cm^{3}$ and with the use of Eqs. (\ref
{17}) we obtain
\begin{equation}\label{31}
\rho _{s}R_{N}^{2}=\frac{3X\left( 3+X\right) }{\left( 2+\Delta _{0}\right)
\left( 1+X\right) ^{2}}, 
\end{equation}
or
\begin{equation}\label{32}
R_{N}=18.891u_{iso}^{1/2}\left( 9-8u_{iso}\right) ^{1/2}\left( 2+\Delta
_{0}\right) ^{-1/2} km,  
\end{equation}
where $R_{N}$ is the radius of the neutron star corresponding to a specific
surface density. For the mass $M_{N}$ and anisotropy parameter $\Delta _{N}$
we find
\begin{equation}\label{33}
M_{N}=25.568u_{iso}^{3/2}\left( 9-8u_{iso}\right) ^{1/2}\left(
2+\Delta _{0}\right) ^{-3/2} M_{\odot},  
\end{equation}
\begin{equation}\label{34}
\Delta _{N}=7.19004\times 10^{35}\Delta _{0}u_{iso}\left( 9-8u_{iso}\right)
^{-1} dyne{\ }cm^{-2}.
\end{equation}

With $\Delta _{0}=0$ we recover the results for $\left(M_{N},R_{N}\right)$ given by
Durgapal and Bannerji \cite{DuBa83}. In Eqs. (\ref{32})-(\ref{34}), for the sake of simplicity,
we have expressed all the quantities in international units, instead of natural units,
by means of the transformations
$\frac{M_{N}}{R_{N}}\rightarrow 8\pi \frac{GM_{N}}{c^{2}R_{N}}$, $\rho
\rightarrow \rho c^{2}$ and $\Delta _{N}\rightarrow \frac{8\pi G}{c^{4}}%
\Delta _{N}$, where $G=6.6732\times 10^{-8}dyne{\ }cm^{2}{\ }g^{-2}$ and $%
c=2.997925\times 10^{10}cm{\ }s^{-1}$. 
 
The variation of the anisotropy parameter $\Delta _{N}$ of the neutron star
as a function of the radius $R_{N}$ and mass $M_{N}$ is represented in Fig. 3.

\newpage

\begin{figure}[h]
\epsfxsize=10cm
\centerline{\epsffile{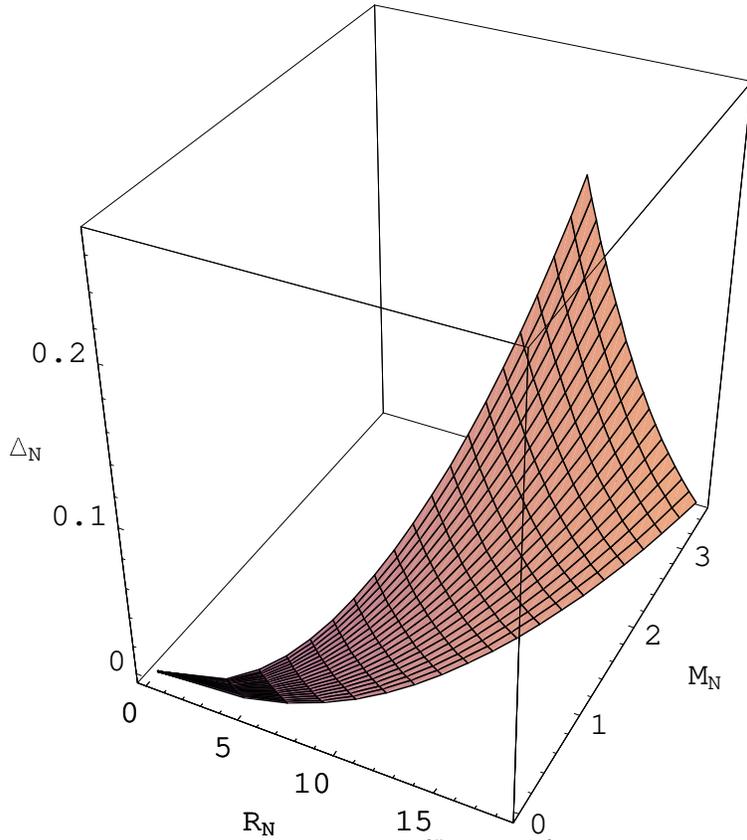}}
\caption{Variation of the anisotropy parameter $\Delta _{N}$  (in units of $10^{35} dynecm^{-2}$) of the neutron star as a
function of the radius $R_{N} (km)$ and  mass $M_{N}$ (in solar mass units)
for $\Delta _{0}\in \left[ 0,0.9\right] $ and $u_{iso}\in \left[ 0.0001,0.28\right] $
.}
\label{FIG3}
\end{figure}

For a particular choice of the equation of state at the center of
the star, $3p_{r0}=3p_{\perp 0}=\rho _{0}$ and with a vanishing anisotropy
parameter, $\Delta _{0}=0$, we obtain the result $u_{iso}=0.2908526$ for a
static isotropic fluid sphere that can be compared with the value of $u_{iso}
$ given in \cite{DuBa83}.

The quantities $\left( \frac{dp_{r}}{%
d\rho }\right) _{r=0}$, $\left( \frac{dp_{\perp }}{d\rho }\right) _{r=0}$ , $%
\left( \frac{dp_{r}}{d\rho }\right) _{r=R}$ and $\left( \frac{dp_{\perp }}{%
d\rho }\right) _{r=R}$ are represented against the anisotropy parameter $%
\Delta _{0}$ in Fig. 4.

\newpage

\begin{figure}[h]
\epsfxsize=10cm
\centerline{\epsffile{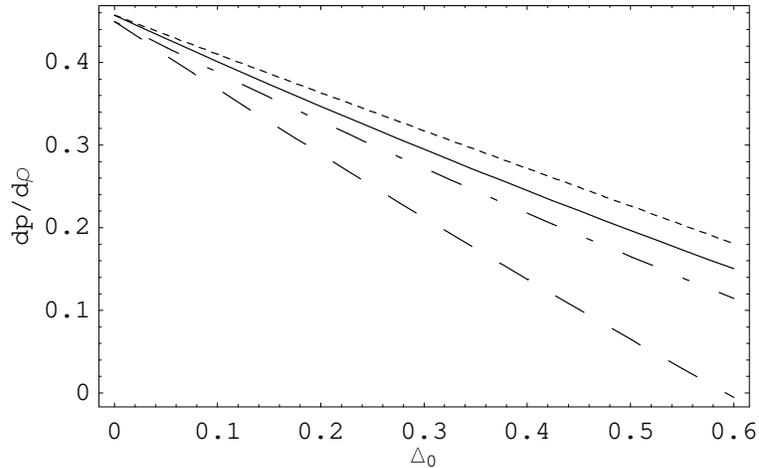}}
\caption{Variations of the radial and tangential speeds of sound $dp/d\rho $
at the vacuum boundary and at the center of the anisotropic fluid sphere as a function of the
anisotropy parameter $\Delta _{0}$ for $u_{iso}=0.24$: 
$\left( dp_{r}/d\rho \right) _{r=R}$ (dotted curve); $\left( dp_{\perp }/d\rho \right)
_{r=R}$ (full curve); $\left( dp_{r}/d\rho \right) _{r=0}$ (dashed-dotted curve); $\left( dp_{\perp }/d\rho
\right) _{r=0}$ (dashed curve).}
\label{FIG4}
\end{figure}

The plots indicate that the necessary and sufficient
criterion for the adiabatic speed of sound to be less than the speed of
light is satisfied by our solution. However, Caporaso and Brecher \cite
{CaBr79} claimed that $dp/d\rho $ does not represent the signal speed. If
therefore this speed exceeds the speed of light, this does not necessary
mean that the fluid is non-causal. But this argument is quite controversial
and not all authors accept it \cite{Gl83}.

\section{Discussions and Final Remarks}

Curvature is described by the tensor field $R_{ijk}^{l}$. It is well known
that if one uses singular behavior of the components of this tensor or its
derivatives as a criterion for singularities, one gets into trouble since the
singular behavior of components could be due to singular behavior of the
coordinates or tetrad basis rather than that of the curvature itself. To
avoid this problem, one should examine the linear and quadratic scalars
formed out of curvature, such as $r_{1}=R$, $r_{2}=R_{ij}R^{ij}$ and $%
r_{3}=R_{ijkl}R^{ijkl}$. 

With the use of the gravitational field equations   (\ref{7})- (\ref{8}) and
of the static line element  (\ref{4}) we obtain the following expressions
for the linear and quadratic scalars of the curvature tensor, given in terms
of the radial pressure, energy density, mass and anisotropy parameter:
\begin{equation}
r_{1}=3p_{r}-\rho +2\Delta ,
\end{equation}
\begin{equation}
r_{2}=\rho ^{2}+3p_{r}^{2}+2\Delta \left( \Delta +2p_{r}\right) ,
\end{equation}
\begin{equation}\label{35}
r_{3}=\left( 2\Delta +\rho +p_{r}-\frac{4m}{r^{3}}\right) ^{2}+2\left( p_{r}+%
\frac{2m}{r^{3}}\right) ^{2}+2\left( \rho -\frac{2m}{r^{3}}\right)
^{2}+4\left( \frac{2m}{r^{3}}\right) ^{2}.  
\end{equation}

When $\Delta =\rho =p_{r}=0$, we recover the invariant $R_{ijkl}R^{ijkl}$ for the
Schwarzschild line element, that is $r_{3}=48m^2/r^6$. 
The variations of $r_{1}$, $r_{2}$ and $r_{3}$ are represented in Fig. 5.

\newpage

\begin{figure}[h]
\epsfxsize=10cm
\centerline{\epsffile{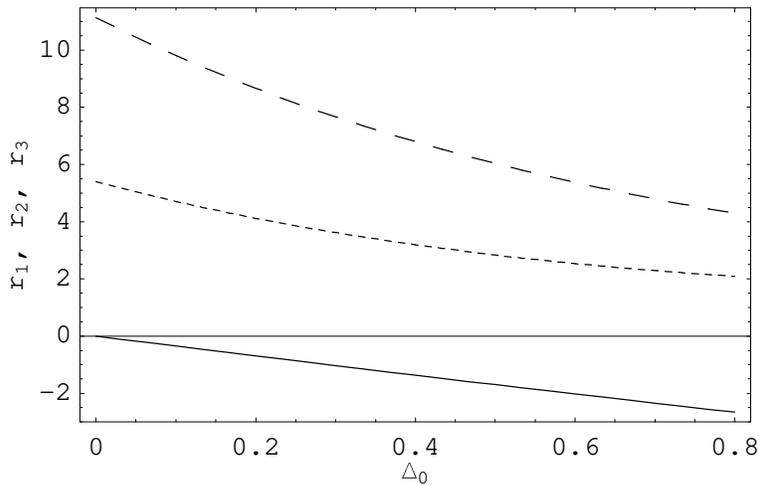}}
\caption{Variation of the curvature scalar $r_{1}$ (solid curve) and of the linear and quadratic
scalars of the curvature tensor $r_{2}$ (dotted curve) and $r_{3}$ (dashed curve) of the static anisotropic fluid
sphere against the anisotropy parameter $\Delta _{0}$ for $u_{iso}=0.2908526$.}
\label{FIG5}
\end{figure}

The scalars $r_{2}$ and $r_{3}$ are finite at the center of the fluid sphere
and monotonically decrease as the anisotropy parameter $\Delta _{0}$
increases.

It is generally held that the trace $T$ of the
energy-momentum tensor must be non-negative. It is also the case that this
trace condition is everywhere fulfilled if it is fulfilled at the center of
the star \cite{Kn88}. 

The purpose of the present paper is to present some exact models of static
anisotropic fluid stars and to investigate their possible astrophysical
relevance. We have extended and generalized to the anisotropic case the
method of obtaining exact solutions for relativistic spheres of Buchdahl 
\cite{Bu59} and Durgapal and Bannerji \cite{DuBa83}. All the solutions we
have obtained are non-singular inside the anisotropic sphere, with finite
values of the density and pressure at the center of the star. Variations of
the physical parameters mass, radius, redshift and adiabatic speed of sound
against the anisotropy parameter have been presented graphically. Our model
can be used to study the interior structure of the anisotropic relativistic
objects because it satisfies all the physical conditions and requirements
(a)-(f).

\section*{Acknowledgments}

We would like to thank Prof. Peter N. Dobson, Jr. for useful discussions and
to the unknown referee whose comments helped us to improve the manuscript.

\end{document}